\documentclass{article}

\usepackage{arxiv}

\usepackage[utf8]{inputenc} 
\usepackage[T1]{fontenc}    
\usepackage{hyperref}       
\usepackage{url}            
\usepackage{booktabs}       
\usepackage{amsfonts}       
\usepackage{nicefrac}       
\usepackage{microtype}      

\usepackage{amsmath, amssymb, amsthm}
\usepackage{algorithm, algorithmic}

\usepackage{mathtools}

\DeclareMathOperator*{\argmin}{arg\,min}

\usepackage{caption}
\usepackage{subcaption}
\usepackage{color}
\usepackage{arydshln}

\newtheorem{theorem}{Theorem}[section]

\newtheorem{lemma}{Lemma}[section]

\newtheorem{corollary}{Corollary}[section]

\usepackage{algorithm}
\usepackage{algorithmic}
\usepackage{amsmath,amssymb}
\usepackage{tikz}
\usetikzlibrary{positioning, arrows.meta}
\allowdisplaybreaks[4]

\title{Fast approximation algorithms for the 1-median problem on real-world large graphs}

\author{
    Keisuke~Ueta\thanks{\texttt{ueta.keisuke.21@shizuoka.ac.jp}.}\\
    Shizuoka University\\
    \And
    Wei~Wu\thanks{\texttt{goi@shizuoka.ac.jp}.}\\
    Shizuoka University\\
    \And
    Mutsunori~Yagiura\thanks{\texttt{yagiura@i.nagoya-u.ac.jp}.}\\
    Nagoya University\\
}

\begin{document}
\maketitle

\begin{abstract}
The 1-median problem (1MP) on undirected weighted graphs seeks to find a facility location minimizing the total weighted distance to all customer nodes. Although the 1MP can be solved exactly by computing the single-source shortest paths from each customer node, such approaches become computationally expensive on large-scale graphs with millions of nodes. In many real-world applications, such as recommendation systems based on large-scale knowledge graphs, the number of nodes (i.e., potential facility locations) is enormous, whereas the number of customer nodes is relatively small and spatially concentrated. In such cases, exhaustive graph exploration is not only inefficient but also unnecessary. Leveraging this observation, we propose three approximation algorithms that reduce computation by terminating Dijkstra’s algorithm early.
We provide theoretical analysis showing that 
one of the proposed algorithms guarantees an approximation ratio of 2, whereas the other two improve this ratio to 1.618. 
We demonstrate that the lower bound of the approximation ratio is 1.2 by presenting a specific instance. Moreover, we show that all proposed algorithms return optimal solutions when the number of customer nodes is less than or equal to three.
Extensive experiments demonstrate that our algorithms significantly outperform baseline exact methods in runtime while maintaining near-optimal accuracy across all tested graph types. Notably, on grid graphs with 10 million nodes, our algorithms obtains all optimal solutions within 1 millisecond, whereas the baseline exact method requires over 70 seconds on average.
\end{abstract}


\section{Introduction}
\label{sec:introduction}
Facility location problems refer to a broad class of problems concerned with determining the optimal placement of facilities to achieve specific objectives. These problems have numerous practical applications in the placement of warehouses, retail stores, fire stations, and other essential infrastructure \cite{owen1998strategic}. The objectives of such problems vary widely, including minimizing total installation costs, minimizing annual operating costs, maximizing service coverage, minimizing average travel time or distance, minimizing maximum travel time or distance, and minimizing the number of facilities to be installed \cite{farahani2010multiple}.

For instance, in the case of locating emergency response facilities such as fire stations, police stations, or hospitals, minimizing the maximum travel time is critical. The key concern is the maximum time it takes to reach any destination after receiving an emergency call. This leads to the so-called center problem. When the number of facilities $p$ to be located is fixed in advance, this is referred to as the $p$-center problem.

On the other hand, transportation costs and service efficiency are related to average travel time or distance. The problem of locating facilities to minimize the average travel time or distance is known as the median problem, which was first proposed by Hakimi in 1964 \cite{hakimi1964optimum}. When the number of facilities is fixed and denoted by $p$, the problem is referred to as the $p$-median problem ($p$MP), which is widely used in applications such as warehouse and store placement.

It has been shown by Kariv and Hakimi \cite{hakimi_1979_treen2p2} that the $p$MP is $\mathcal{NP}$-hard for $p \ge 2$. Revelle and Swain \cite{revelle1970central} formulated the $p$MP as a mixed-integer programming problem. Various methods have been proposed to solve the $p$MP, including Lagrangian relaxation \cite{BEASLEY1985270,doi:10.1287/opre.25.4.709} and branch-and-price algorithms \cite{senne2005branch}. In particular, Lagrangian relaxation methods using subgradient optimization have been shown to be effective for solving large-scale instances of the $p$MP \cite{daskin2015p}.

Many researchers have also studied special cases of the $p$MP. For example, when the graph is a tree, an algorithm with a time complexity of $\mathcal{O}(pn^2)$ is known \cite{tamir1996pn2}, where $n$ is the number of nodes in the graph. Furthermore, when the graph is a tree and $p=1$, a linear-time algorithm exists \cite{goldman1971optimal}.
Additionally, many heuristic methods have been proposed for solving the $p$MP in practical applications \cite{alp2003efficient, daskin1997network, teitz1968heuristic}.

Despite this extensive body of work, the 1-median problem (1MP), in which only one facility is to be located, has received comparatively little attention, because it can be solved in polynomial time.
However, when the input graph is extremely large (e.g., a knowledge graph with millions of nodes), even polynomial-time algorithms may result in prohibitively long computation times. Moreover, precomputing and storing all-pairs shortest distances requires $\mathcal{O}(n^2)$ space, which is often impractical. A motivating example of the 1MP considered in this study is the recommendation of the next node (i.e., a facility node) to a user based on a knowledge graph and a set of customer nodes (e.g., nodes of interest inferred from user history). In real-time applications such as recommendation systems, it becomes practically difficult to respond within a short time window using exact methods.

In this study, we aim to design algorithms that reduce computation time for the 1MP by appropriately narrowing the search space. Our focus is on large-scale graphs with millions of nodes (potential facility locations), where the number of customer nodes is relatively small and they are spatially concentrated. 
We propose three approximation algorithms and show that one of them guarantees an approximation ratio of 2, whereas the other two improve this ratio to 1.618. We validate the effectiveness of these algorithms by comparing them with exact methods based on computing single-source shortest paths from each customer node. In particular, we confirm that our methods can obtain optimal solutions for all tested graphs with 10 million nodes, demonstrating their suitability for commercial applications.

\section{Problem description}
\label{sec:problem-description}
We are given a connected, undirected graph $G = (V, E)$ with a node set $V = \{1, 2, \ldots, n\}$ and an edge set $E$, where each edge $\{i, j\} \in E$ is associated with a non-negative cost (distance) $c_{ij}$ ($= c_{ji}$). We are also given a set of customer nodes $M \subseteq V$ with $|M| = m$, where each customer node $j \in M$ is assigned a weight $w_j$. The 1-median problem (1MP) considered in this study is to seek a node (facility location) that minimizes the total weighted shortest distances from all customer nodes. Without loss of generality, we assume $M = \{1, 2, \ldots, m\}$.

Some previous studies define the 1MP with $M = V$, which is clearly a special case of the definition described in this study. Moreover, the two problem definitions are equivalent, as our setting can be derived from theirs by assigning a weight $w_j = 0$ to all non-customer nodes.


For convenience, let $c_{ji}^{\mathrm{(sp)}}$ denote the shortest-path distance from node $j$ to node $i$ in the graph. If node $i \in V$ is chosen as the facility location, we define its evaluation value as $z(i)=\sum_{j\in M}w_jc_{ji}^\mathrm{(sp)}$. Then, the 1MP can be formulated as:
\begin{align*}
    \min_{i\in V}z(i)=\min_{i\in V}\sum_{j\in M} w_j c_{ji}^{\mathrm{(sp)}}.
\end{align*}

\section{Exact Method} \label{sec:exact}
As a baseline exact method for the 1MP, we can determine the optimal facility location computing the single-source shortest paths from each customer node.


An exact approach based on Dijkstra’s algorithm is presented in Algorithm~\ref{argo:1}. Algorithm~\ref{argo:1} executes Dijkstra’s algorithm $m$ times, once for each customer node. The computational complexity of a single execution depends on the data structure used for the priority queue. When a binary heap is used, the complexity of Algorithm~\ref{argo:1} is $\mathcal{O}(m(|E| + n)\log n)$; with a Fibonacci heap, it improves to $\mathcal{O}(m|E| + mn\log n)$.

\begin{algorithm}[htb]
\caption{An exact method using Dijkstra's algorithm.}\label{argo:1}
\begin{algorithmic}[1]
    \FORALL{$j \in M$}
        \STATE Run Dijkstra's algorithm from source node $j$ to compute $c_{ji}^{\mathrm{(sp)}}$ for all $i \in V$.
    \ENDFOR
    \FORALL{$i \in V$}
        \STATE Compute $z(i) \gets \sum_{j \in M} w_j c_{ji}^{\mathrm{(sp)}}$.
    \ENDFOR
    \RETURN $\min_{i \in V} z(i)$.
\end{algorithmic}
\end{algorithm}

\section{Proposed Methods} \label{sec:proposed}
A key characteristic of the graphs arising in our target applications is that, whereas the overall graph is large, the induced subgraph that minimally connects all customer nodes is relatively small. As a result, nodes far from the customer nodes are unlikely to be an optimal facility location. This is especially true when customer nodes are densely clustered, in which case Dijkstra's algorithm from each customer often reaches the true optimal facility node early in its execution.

Based on this observation, we propose three approximation algorithms that terminate Dijkstra's algorithm early, rather than running it to completion: 
\begin{itemize} 
\item Truncated Dijkstra algorithm with selective aggregation (TDA-SA), 
\item Truncated Dijkstra algorithm with nearest-neighbor approximation (TDA-NNA), 
\item Truncated Dijkstra algorithm with shortest-path approximation (TDA-SPA). 
\end{itemize}







In all three algorithms, for each customer node, Dijkstra’s algorithm is terminated once the shortest paths to all other customer nodes have been determined. The pseudocode for TDA-SA, TDA-NNA, and TDA-SPA are shown in Algorithms~\ref{algo:teian_1}, \ref{algo:teian_2}, and \ref{algo:teian_3}, respectively.
We describe the differences among the three algorithms below.

TDA-SA considers only nodes for which the shortest-path distances from \emph{all} customer nodes have been determined as candidates for the facility location. In this paper, we say that a node $i$ is \emph{determined} from a customer node $j$ if the shortest-path distance from $j$ to $i$ has been finalized during the execution of Dijkstra’s algorithm from source node $j$. Because all necessary distances to such candidate nodes are available, their objective value $z(i)$ can be computed exactly.

TDA-NNA expands the candidate set to include nodes for which the shortest-path distance from \emph{at least one} customer node has been determined. For a candidate node $i$, if the shortest-path distance from a customer node $j$ has not yet been determined, we approximate $c^{\mathrm{(sp)}}_{ji}$ using the nearest known customer node $j'$ to $i$. That is, the distance is approximated as:
\begin{align*}
c^{\mathrm{(sp)}}_{ji}\approx c^{\mathrm{(nna)}}_{ji} = c^{\mathrm{(sp)}}_{jj'} + c^{\mathrm{(sp)}}_{j'i},
\end{align*}
where both $c^{\mathrm{(sp)}}_{jj'}$ and $c^{\mathrm{(sp)}}_{j'i}$ have already been determined.

TDA-SPA uses the same candidate nodes as TDA-NNA, but it obtains a solution that is always at least as good, and potentially better. If the shortest-path distance from customer node $j$ to candidate node $i$ is not known, TDA-SPA approximates it by choosing the minimum possible sum of known distances via any intermediate customer node $j'$:
\begin{align*}
c^{\mathrm{(sp)}}_{ji}\approx c^{\mathrm{(spa)}}_{ji} = \min_{j':\ c^\mathrm{(sp)}_{j'i} \text{ is determined}}\left\{c^{\mathrm{(sp)}}_{jj'} + c^{\mathrm{(sp)}}_{j'i}\right\}.
\end{align*}
TDA-SPA guarantees an objective value that is no worse than that of TDA-NNA. However, it incurs additional computational cost. After running Dijkstra's algorithm $m$ times (once per customer), the additional approximation phase (Steps 6-10 in Algorithm \ref{algo:teian_3}) takes $\mathcal{O}(m^2 n)$ time, which may become a bottleneck for some instances.

Note that TDA-NNA and TDA-SPA select the facility location based on approximate evaluations. In practice, once the final facility location $i$ is determined, the exact objective value $z(i)$ can be computed by running Dijkstra's algorithm from node $i$ until the shortest paths to all customer nodes are obtained.

\begin{algorithm}[htb]
\caption{Truncated Dijkstra algorithm with selective aggregation (TDA-SA).}\label{algo:teian_1}
\begin{algorithmic}[1]
\STATE Set $c_{ji}^{\mathrm{(alg)}} \gets \infty$ for all $j \in M$ and $i \in V$.
\FORALL{$j \in M$}
\STATE Run Dijkstra's algorithm from node $j$ to update $c_{ji}^{\mathrm{(alg)}}$ until the shortest-path distances to all nodes in $M$ have been determined.
\ENDFOR
    \STATE Let $V' \gets \{i \in V \mid \forall j\in M, \text{the shortest-path distances from $j$ to $i$ has been determined}\}$.
    \FORALL{$i \in V'$}
        \STATE Compute $z_{\text{sa}}(i) \gets \sum_{j \in M} w_j \, c_{ji}^{\mathrm{(alg)}}$.
    \ENDFOR
    \RETURN $\min_{i \in V'} z_{\text{sa}}(i)$.
\end{algorithmic}
\end{algorithm}

\begin{algorithm}[htb]
\caption{Truncated Dijkstra algorithm with nearest-neighbor approximation (TDA-NNA).}\label{algo:teian_2}
\begin{algorithmic}[1]
\STATE Set $c_{ji}^{\mathrm{(alg)}} \gets \infty$ for all $j \in M$ and $i \in V$.
\FORALL{$j \in M$}
\STATE\label{step:termination} Run Dijkstra's algorithm from node $j$ to update $c_{ji}^{\mathrm{(alg)}}$ until the shortest-path distances to all nodes in $M$ have been determined.
\ENDFOR
\STATE $V''\gets \{i\in V\mid \exists j\in M, \text{the shortest-path distance from $j$ to $i$ has been determined}\}$.
\FORALL{$i \in V''$}
\STATE Find $j'$, the customer node closest to $j$, that is, $j'\gets \argmin_{j \in M} c_{ji}^{\mathrm{(alg)}}$.
\FORALL{$j \in M$}
\STATE $c_{ji}^{\mathrm{(nna)}} \gets \min\left\{c_{jj'}^\mathrm{(alg)} + c_{j'i}^\mathrm{(alg)},c_{ji}^\mathrm{(alg)}\right\}.$
\ENDFOR
\ENDFOR
    \FORALL{$i \in V''$}
        \STATE Compute $z_{\mathrm{nna}}(i) \gets \sum_{j \in M} w_j c_{ji}^{\mathrm{(nna)}}$.   
 \ENDFOR
\RETURN $\min_{i \in V''} z_{\mathrm{nna}}(i)$.
\end{algorithmic}
\end{algorithm}

\begin{algorithm}[htb]
\caption{Truncated Dijkstra algorithm with shortest-path approximation (TDA-SPA).}\label{algo:teian_3}
\begin{algorithmic}[1]
\STATE Set $c_{ji}^{\mathrm{(alg)}} \gets \infty$ for all $j \in M$ and $i \in V$.
\FORALL{$j \in M$}
\STATE Run Dijkstra's algorithm from node $j$ to update $c_{ji}^{\mathrm{(alg)}}$ until the shortest-path distances to all nodes in $M$ have been determined.
\ENDFOR
\STATE $V''\gets \{i\in V\mid \exists j\in M, \text{the shortest-path distance from $j$ to $i$ has been determined}\}$.
\FORALL{$i \in V''$}
\FORALL{$j \in M$}
\STATE $c_{ji}^{\mathrm{(spa)}} \gets \min_{j' \in M}\left\{c_{jj'}^\mathrm{(alg)} + c_{j'i}^\mathrm{(alg)},c_{ji}^\mathrm{(alg)}\right\}.$
\ENDFOR
\ENDFOR
    \FORALL{$j \in V''$}
        \STATE Compute $z_{\text{spa}}(j) \gets \sum_{j \in M}w_j c_{ji}^{\mathrm{(spa)}}$.   
 \ENDFOR
\RETURN $\min_{j \in V''} z_{\text{spa}}(j)$.
\end{algorithmic}
\end{algorithm}

\color{black}

\section{Approximation Accuracy Analysis}

In this section, we theoretically analyze the solution quality obtained by the three proposed methods introduced in Section~\ref{sec:proposed}.

Recall that $z_\mathrm{sa}(j)$, $z_\mathrm{nna}(j)$, and $z_\mathrm{spa}(j)$ denote the evaluation values computed by Algorithms~\ref{algo:teian_1}, \ref{algo:teian_2}, and \ref{algo:teian_3}, respectively.

Due to the termination condition of Dijkstra's algorithm described in Step~\ref{step:termination} of Algorithms~\ref{algo:teian_1}–\ref{algo:teian_3}, the following lemma holds for all three proposed methods:
\begin{lemma}\label{lem:opt-M}
For every $j\in M$, we have $z(j)=z_\mathrm{sa}(j)=z_\mathrm{nna}(j)=z_\mathrm{spa}(j)$.
\end{lemma}

We first prove that when the number of customer nodes $m$ is less than or equal to 3, all three proposed methods return an optimal solution.
\begin{theorem}\label{the:allopt}
When $m \leq 3$, the solutions obtained by all three proposed methods are optimal.
\end{theorem}
\begin{proof}
The cases $m = 1$ and $m = 2$ are trivial. We consider the case $m = 3$, that is, $M = \{1, 2, 3\}$.

Let $v^* \in V$ be an arbitrary optimal facility location. If the shortest-path distances from all customer nodes to $v^*$ have been determined by the proposed methods, then $v^*$ is considered as a candidate, and its objective value $z(v^*)$ is exactly computed. Thus, the optimal solution is returned.

We now consider the non-trivial case where the shortest-path distance from at least one customer node to the optimal facility location $v^*$ has not been determined. Without loss of generality, assume the following:
\begin{enumerate}
\item The shortest-path distance $c_{1v^*}^{\mathrm{(sp)}}$ has not been determined.
\item The customer weights satisfy $w_2 \geq w_3$.
\end{enumerate}

From the first assumption, and because Dijkstra's algorithm terminates when shortest-path distances to all customer nodes are determined, we must have:
\begin{align}\label{3pt_proof:1}
c_{12}^{\mathrm{(sp)}} \leq c_{1v^*}^{\mathrm{(sp)}}.
\end{align}
Combining inequality~\eqref{3pt_proof:1} with the triangle inequality $c_{23}^\mathrm{(sp)} \leq c_{2v^*}^\mathrm{(sp)} + c_{3v^*}^\mathrm{(sp)}$, and using the second assumption $w_2 \geq w_3$, we obtain:
\begin{align*}
    z(2) &= w_1 c_{12}^{\mathrm{(sp)}} + w_3 c_{32}^{\mathrm{(sp)}}\leq w_1 c_{12}^{\mathrm{(sp)}} + \left(w_3 c_{2v^*}^{\mathrm{(sp)}} + w_3 c_{3v^*}^{\mathrm{(sp)}}\right) \\
    &\leq w_1 c_{1v^*}^{\mathrm{(sp)}} + w_2 c_{2v^*}^{\mathrm{(sp)}} + w_3 c_{3v^*}^{\mathrm{(sp)}} = z(v^*).
\end{align*}

From Lemma~\ref{lem:opt-M}, we have:
\begin{align*}
z_\mathrm{sa}(2)=z_\mathrm{nna}(2)=z_\mathrm{spa}(2)=z(2)\le z(v^*),
\end{align*}
which implies that node 2 is at least as good as $v^*$ in terms of the objective value.

Hence, all three proposed methods return an optimal solution when $m \leq 3$.
\end{proof}

Next, we derive a tight approximation ratio of TDA-SA for the \emph{unweighted 1MP}, where $w_j = 1$ for every $j \in M$.


\begin{theorem}\label{thm:>=3}
For the unweighted 1MP with $m \ge 4$, the approximation ratio of TDA-SA is $\left(2 - \frac{4}{m+1}\right)$.
\end{theorem}

\begin{proof}
Let $v^*$ be an arbitrary optimal solution, and let $v_\mathrm{sa}$ be the solution obtained by TDA-SA. If the shortest-path distances from all customer nodes to $v^*$ are determined (i.e., $z_\mathrm{sa}(v^*) = z(v^*)$), then $v^*$ is included in the candidate set, and the algorithm returns the optimal solution. In that case, the theorem holds trivially.

Otherwise, without loss of generality, we assume that the shortest-path distance from customer node 1 to $v^*$ was not determined. Because TDA-SA terminates Dijkstra’s algorithm once all shortest-path distances to customer nodes are computed, we have:
\begin{align}
c^\mathrm{(sp)}_{1v^*} \ge c^\mathrm{(sp)}_{1i} &\quad \forall i \in M. \label{eq:ishort}
\end{align}
From \eqref{eq:ishort} and the definition of $v_\mathrm{sa}$, it follows that:
\begin{align}
\label{eq:p1}c^\mathrm{(sp)}_{1v^*} \ge \frac{1}{m-1} \sum_{i=2}^{m} c^\mathrm{(sp)}_{1i}
= \frac{1}{m-1} z(1) \ge \frac{1}{m-1} z(v_\mathrm{sa}).
\end{align}
Using the triangle inequality and the fact that Dijkstra’s algorithm terminates once all customer nodes are reached, we can derive the following:
\begin{align}
\nonumber (m-1) \cdot z(v_\mathrm{sa}) &\le \sum_{i=2}^{m} z(i) = \sum_{i=2}^{m} c^\mathrm{(sp)}_{1i} + 2\sum_{i=2}^{m-1} \sum_{j=i+1}^{m} c^\mathrm{(sp)}_{ij}\\
\nonumber &\le \sum_{i=2}^{m} c^\mathrm{(sp)}_{1i} + 2\sum_{i=2}^{m-1} \sum_{j=i+1}^{m} \left(c^\mathrm{(sp)}_{iv^*} + c^\mathrm{(sp)}_{jv^*}\right)\\
\label{eq:p2}&= \sum_{i=2}^{m} c^\mathrm{(sp)}_{1i} + 2(m-2) \sum_{i=2}^{m} c^\mathrm{(sp)}_{iv^*}.
\end{align}
Combining inequalities \eqref{eq:p1} and \eqref{eq:p2}, we obtain:
\begin{align*}
z(v^*) &= c^\mathrm{(sp)}_{1v^*} + \sum_{i=2}^{m} c^\mathrm{(sp)}_{iv^*}\\
&= \frac{m-3}{2(m-2)} c^\mathrm{(sp)}_{1v^*} + \frac{m-1}{2(m-2)} c^\mathrm{(sp)}_{1v^*} + \sum_{i=2}^{m} c^\mathrm{(sp)}_{iv^*}\\
&\ge \frac{m-3}{2(m-2)} c^\mathrm{(sp)}_{1v^*} + \frac{1}{2(m-2)} \sum_{i=2}^{m} c^\mathrm{(sp)}_{1i} + \sum_{i=2}^{m} c^\mathrm{(sp)}_{iv^*}\\
&\ge \frac{m-3}{2(m-1)(m-2)} z(v_\mathrm{sa}) + \frac{m-1}{2(m-2)} z(v_\mathrm{sa})\\
&= \frac{m+1}{2(m-1)} z(v_\mathrm{sa}).
\end{align*}
Therefore, the approximation ratio is bounded as:
\begin{align*}
\frac{z_\mathrm{sa}(v_\mathrm{sa})}{z(v^*)} = \frac{z(v_\mathrm{sa})}{z(v^*)} = 2 - \frac{4}{m+1}.
\end{align*}
\end{proof}

Recall that $V'$ is the candidate set used in TDA-SA, and $V''$ is the candidate set used in both TDA-NNA and TDA-SPA, with $V' \subseteq V''$.
Because for every $i \in V'$ (i.e., $i \in V''$), it holds that $z_\mathrm{nna}(i) \leq z_\mathrm{sa}(i)$ and $z_\mathrm{spa}(i) \leq z_\mathrm{sa}(i)$, the same approximation ratio also holds for both TDA-NNA and TDA-SPA.

\begin{corollary}\label{cor:also_nnaspa}
For the unweighted 1MP with $m \ge 4$, the approximation ratios of TDA-NNA and TDA-SPA are also given by $\left( 2 - \frac{4}{m+1} \right)$.
\end{corollary}

We now show that the approximation ratio given in Theorem~\ref{thm:>=3} is tight for TDA-SA.

\begin{lemma}\label{lem:k=4}
For the unweighted 1MP, the approximation ratio of TDA-SA stated in Theorem~\ref{thm:>=3} is tight when $m \ge 4$.
\end{lemma}
\begin{proof}
To show that the ratio $\left(2 - \frac{4}{m+1} \right)$ in Theorem~\ref{thm:>=3} is tight, we construct the following instance.

Let $G$ be a complete graph with $m+1$ vertices, consisting of $m$ customer nodes and one non-customer node. Define the cost $c_{ij}$ as follows:
\begin{align*}
c_{ij}=\begin{cases}
2 &\text{if $i,j\in \{1,2,\ldots,m\}$ and $i\ne j$}\\
2+\epsilon &\text{if $i=1$ and $j=m+1$}\\
1 &\text{if $i\in \{2,\ldots,m\}$ and $j=m+1$,}
\end{cases}
\end{align*}
where $\epsilon > 0$ is an arbitrarily small positive value.

For this instance, the evaluation value of TDA-SA is:
\begin{align*}
&z_\mathrm{sa}(i) = 2(m-1) &\forall i\in \{1,2,\ldots,m\}\\
&z_\mathrm{sa}(m+1) = \infty
\end{align*}
On the other hand, the optimal 
solution is achieved by placing the facility at node $m+1$, with the total cost:
\begin{align*}
z(m+1) = (m - 1) \cdot 1 + (2 + \epsilon) = m + 1 + \epsilon.
\end{align*}

Thus, the approximation ratio becomes:
\begin{align*}
\frac{\min_{i\in V}z_\mathrm{sa}(i)}{z(m+1)} = \frac{2(m-1)}{m+1+\epsilon} = 2 - \frac{4+2\epsilon}{m+1+\epsilon}
\end{align*}
which approaches $\left(2 - \frac{4}{m+1} \right)$ as $\epsilon \to 0$. This confirms that the bound in Theorem \ref{thm:>=3} is tight.
\end{proof}

In Theorem~\ref{thm:>=3} and Corollary~\ref{cor:also_nnaspa}, we obtained theoretical results for the unweighted 1MP. Next, we show that these results can be extended to the (weighted) 1MP.

\begin{theorem}\label{theorem:2kinzi}
For the (weighted) 1MP, the approximation ratio of the three proposed methods is $2$.
\end{theorem}

\begin{proof}
We prove this theorem by showing that node weights do not affect the approximation ratio analysis presented in Theorem~\ref{thm:>=3} and Corollary~\ref{cor:also_nnaspa}.

Because $w_j\in \mathbb{R}_{\ge 0}$, we can convert them to integers via common denominators does not affect the optimal solution.  
Without loss of generality, assume the weights $w_j$ for each customer node $j$ are integers.  
Given a weighted instance, each customer node $j$ with weight $w_j$ can replaced by $w_j$ unit-weight customer nodes that are interconnected with edges of cost zero. Each of these new nodes is then connected to the original neighbors of node $j$ using the same edge costs as in the original graph.
This transformation preserves all shortest-path distances relevant to the 1MP objective, effectively reducing the weighted instance to an equivalent unweighted one.
Therefore, node weights do not affect the constant term (value 2) in the approximation ratio established in Theorem~\ref{thm:>=3} and Corollary~\ref{cor:also_nnaspa}.
\end{proof}

Before we show that the approximation ratios of TDA-NNA and TDA-SPA can be further improved, we first present a lemma that will be used in the subsequent analysis.

\begin{lemma}\label{lem:0notcalc}
For the unweight 1MP, there exists an optimal node $v^*$ for which the shortest paths from at least two customer nodes are determined.
\begin{proof}
First, consider any node $v$ for which the shortest paths from all customer nodes are not determined.  
Then, for each customer node $j \in M$, the following inequality holds:
\begin{align*}
c^\mathrm{(sp)}_{jv} \ge \frac{1}{m-1}\sum_{i\in M\setminus\{j\}}c^\mathrm{(sp)}_{ji}= \frac{1}{m-1}z(i)\ge \frac{1}{m-1}z(v^*) \quad \forall j\in M.
\end{align*}
Therefore, the evaluation value of node $v$ satisfies:
\begin{align}
z(v) = \sum_{j\in M} c^\mathrm{(sp)}_{jv} \ge \frac{m}{m-1} z(v^*) > z(v^*). \label{0notcalc_2}
\end{align}
Inequality \eqref{0notcalc_2} implies that such a node $v$ cannot be an optimal solution.

Next, consider a node $v'$ for which the shortest-path distance is determined from only one customer node $k$.  
In this case, we have:
\begin{align}
z(v') \ge z(k)\ge z(v^*).
\end{align}
Because node $v'$ is dominated by node $k$, from which all shortest-path distances are known, it cannot be strictly better than $k$.

Therefore, there exists an optimal node $v^*$ from which the shortest paths from at least two customer nodes are determined.
\end{proof}
\end{lemma}

Lemma~\ref{lem:0notcalc} can also be utilized to design an exact algorithm.  
After executing the truncated Dijkstra procedures (Steps~1--4 in Algorithms~\ref{algo:teian_1}--\ref{algo:teian_3}), we can identify all nodes for which the shortest paths from at least two customer nodes have been determined.  
We then continue the truncated Dijkstra process until the objective values of these identified nodes are computed exactly.  
Among these candidates, the node with the smallest objective value is guaranteed to be optimal for the unweighted 1MP.  
For the (weighted) 1MP, the set of candidate nodes should be extended to include all nodes for which the shortest paths from at least one customer nodes have been determined.

Using Lemma~\ref{lem:0notcalc}, we are now ready to present the main theoretical result of this study.

\begin{theorem}
For the unweighted 1MP with $m\ge 4$, the approximation ratio of TDA-NNA and TDA-SPA is at most $1.618$.
\end{theorem}

\begin{proof}
Because $z_{\mathrm{nna}}(i) \ge z_{\mathrm{spa}}(i)$ for all $i \in V''$, TDA-SPA always achieves a solution that is at least as good as that of TDA-NNA. Thus, we focus on analyzing TDA-NNA.

Let $v^*$ be an optimal node, and let $v_\mathrm{nna}$ be the node obtained by TDA-NNA.  
If all shortest-path distances from all customer nodes to $v^*$ are determined, that is, $z_\mathrm{nna}(v^*)=z(v^*)$, then the optimal solution is obtained, and the theorem holds trivially.

Otherwise, by Lemma \ref{lem:0notcalc}, we can assume that at least two customer nodes have finalized their shortest-path distances to $v^*$.  
Without loss of generality, assume that the shortest paths from customer nodes $1,2,\ldots,k$ (where $1\le k\le m-2$) are not determined.  
Let $j'$ be the customer node closest to $v^*$ among the remaining customer nodes $\{k+1, k+2, \ldots,m\}$:
\begin{align*}
j'=\argmin_{j=k+1}^m c^\mathrm{(sp)}_{jv^*}.
\end{align*}
Because $c^{\mathrm{(sp)}}_{ij} \le c^{\mathrm{(sp)}}_{iv^*}$ hold for $i \in \{1,2, \ldots,k\}$ and $j\in\{k+1,k+2,\ldots,m\}$, we have:
\begin{align}\label{bd:1}
\nonumber z_\mathrm{nna}(v_\mathrm{nna})&\le z_\mathrm{nna}(v^*)=\sum_{i=k+1}^m c^\mathrm{(sp)}_{iv^*} + \sum_{i=1}^k \left\{c^\mathrm{(sp)}_{ij'}+c^\mathrm{(sp)}_{j'v^*}\right\}\\
\nonumber&\le \sum_{i=k+1}^m c^\mathrm{(sp)}_{iv^*} + \sum_{i=1}^k c^\mathrm{(sp)}_{iv^*} + kc^\mathrm{(sp)}_{j'v^*}\\
&\le \sum_{i=1}^m c^\mathrm{(sp)}_{iv^*} + \frac{k}{m}z(v^*)=\frac{m+k}{m}z(v^*)
\end{align}
On the other hand, for $i\in \{1,2,\ldots,k\}$, we have:
\begin{align*}
&c^\mathrm{(sp)}_{iv^*} \ge c^\mathrm{(sp)}_{ij} &\forall j\in M, j\ne i.
\end{align*}
Hence,
\begin{align}\label{eq:ele}
c^\mathrm{(sp)}_{iv^*} \ge \frac{1}{m-1}\sum_{j\in M: j\ne i}c^\mathrm{(sp)}_{ij}=\frac{1}{m-1}z(i) \ge \frac{1}{m-1}z_\mathrm{nna}(v_\mathrm{nna}),
\end{align}
which implies:
\begin{align}\label{bd:2}
z(v^*)=\sum_{i=1}^m c^\mathrm{(sp)}_{iv^*}\ge \sum_{i=1}^k c^\mathrm{(sp)}_{iv^*} \ge \frac{k}{m-1}z_\mathrm{nna}(v_\mathrm{nna}).
\end{align}
Combining both bounds in \eqref{bd:1} and \eqref{bd:2} yields:
\begin{align*}
\frac{z_\mathrm{nna}(v_\mathrm{nna})}{z(v^*)}\le\min\left\{\frac{m+k}{m}, \frac{m-1}{k}\right\}\le\min\left\{\frac{m+k}{m}, \frac{m}{k}\right\}.
\end{align*}
Let $k = \alpha m$ where $\alpha \in (0,1)$, the worst-case ratio is:
\begin{align*}
\frac{z_\mathrm{nna}(v_\mathrm{nna})}{z(v^*)} \le \min\left\{1 + \alpha, \frac{1}{\alpha} \right\} \le \frac{1+\sqrt{5}}{2} \approx 1.618.
\end{align*}
\end{proof}

Using the same technique as in the proof of Theorem~\ref{theorem:2kinzi}, the approximation ratio of $1.618$ also holds for the weighted case.

\begin{corollary}\label{cor:16kinzi-weighted}
For the (weighted) 1MP, the approximation ratio of both TDA-NNA and TDA-SPA is also at most $\frac{1+\sqrt{5}}{2} \approx 1.618$.
\end{corollary}

Finally, we present a valid lower bound on the approximation ratio of TDA-NNA and TDA-SPA.


\begin{theorem}\label{theorem:12kinzi}
The approximation ratio of TDA-NNA and TDA-SPA has a lower bound of $1.2$.
\end{theorem}
\begin{proof}
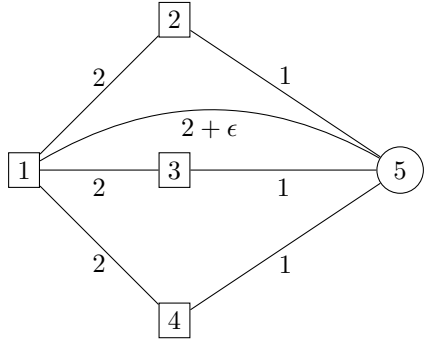
\begin{figure}[htbp]
    \centering
\begin{tikzpicture}
  \node[draw, rectangle] (A) at (-2, 0) {$1$};
  \node[draw, rectangle] (B) at (0, 2) {$2$};
  \node[draw, rectangle] (C) at (0, 0) {$3$};
  \node[draw, rectangle] (D) at (0, -2) {$4$};
  \node[draw, circle] (E) at (3, 0) {$5$};
  
  \draw[-] (A) -- node[pos=0.5, above] {$2$} (B);
  \draw[-] (A) -- node[pos=0.5, below] {$2$} (C);
  \draw[-] (A) -- node[pos=0.5, below] {$2$} (D);
  \draw[-] (B) -- node[pos=0.5, above] {$1$} (E);
  \draw[-] (C) -- node[pos=0.5, below] {$1$} (E);
  \draw[-] (D) -- node[pos=0.5, below] {$1$} (E);
  \draw[-] (A) to[bend left=30] node[pos=0.5, below] {$2+\epsilon$} (E);
  
\end{tikzpicture}
\caption{An example instance with an approximation ratio of $1.2$.}
\label{fig:hanrei}
\end{figure}
Figure~\ref{fig:hanrei} illustrates an instance that attains this lower bound.  
In this instance, we have $n=5$ nodes and $m=4$ customer nodes, each with weight $1$.  
Let $\epsilon > 0$ be an arbitrarily small positive value.  
The optimal facility location is node 5, with an optimal objective value of $5 + \epsilon$.  

However, in both TDA-NNA and TDA-SPA, the shortest-path distance from node 1 to node 5 is approximated by:
\begin{align*}
c_{12} + c_{25} = 3.
\end{align*}
and the objective value computed for each candidate node is 6.  
Thus, the approximation ratio for this instance is: $6 / (5 + \epsilon) \approx 1.2$.
\end{proof}




\section{Computational Results}
In this section, we conduct computational experiments on 1MP using the algorithms introduced in Sections~\ref{sec:exact} and \ref{sec:proposed}, and discuss the obtained results.

\subsection{Computational Environment and Instance Generation}
The exact method (Algorithm~\ref{argo:1}) and the three proposed algorithms (Algorithms~\ref{algo:teian_1}, \ref{algo:teian_2}, \ref{algo:teian_3}) were implemented in C++.
All computational experiments were carried out on a PC equipped with a Xeon E-2286G (4.0~GHz) and 64~GB of memory.

We generated and used the following six types of graphs in the experiments:
\begin{itemize}
    \item \textbf{RRU} (random graph with random source selection and uniform vertex weights): A random graph in which the number of edges $|E|$ is uniformly sampled from $[n-1,\frac{n(n-1)}{2}]$. Initially, a spanning tree with $n-1$ edges is constructed to ensure connectivity, and the remaining edges are added afterward. Edge weights are chosen uniformly at random from $[0,1)$. Customer nodes are selected uniformly at random from $V$, with $m$ nodes chosen, each assigned weight $1$.
    
    \item \textbf{RRW} (random graph with random source selection and weighted vertices): Similar to RRU, but customer node weights are drawn uniformly at random from $[0,1)$ instead of being fixed to 1.
    
    \item \textbf{RNU} (random graph with neighbor-restricted source selection and uniform vertex weights): A random graph with $|E|=4n$, initially constructing a spanning tree to ensure connectivity. Edge weights are drawn uniformly from $[0,1)$. A single node is selected randomly as the source, from which a breadth-first search (BFS) is performed.  Among the $\max\left\{2m,\lfloor\log_2n\rfloor\right\}$ neighboring nodes, $m$ are randomly chosen as customer nodes, each with weight 1.
    
    \item \textbf{RDU} (random graph with distance-restricted source selection and uniform vertex weights): Same as RNU, except that Dijkstra’s algorithm is used instead of BFS to identify the  $\max\left\{2m,\lfloor\log_2n\rfloor\right\}$ nearest neighbors, from which $m$ customers are selected.
    
    \item \textbf{GNU} (grid graph with neighbor-restricted source selection and uniform vertex weights): A grid graph with edge weights uniformly drawn from $[0,1)$. A random source node is selected, and BFS is performed to select $m$ customer nodes from the $\max\left\{2m,\lfloor\log_2n\rfloor\right\}$ neighbors, each with weight 1.
    
    \item \textbf{GDU} (grid graph with distance-restricted source selection and uniform vertex weights): Similar to GNU, but customer nodes are selected using Dijkstra’s algorithm instead of BFS.
\end{itemize}

The instance sets used in the experiments are as follows:
\begin{itemize}
    \item \textbf{SMALL}: $n=50$, $m\in \{1,2,\ldots,n\}$, graph types $t\in \{\mathrm{RRU},\mathrm{RRW}\}$. For each $(n,m,t)$ combination, 50,000 instances were generated, totaling 5,000,000 instances.
    \item \textbf{LARGE}: $n\in \{10^4,10^5,10^6,10^7\}$, $m\in \{2,8,32\}$, graph types $t\in \{\mathrm{RNU},\mathrm{RDU},\mathrm{GNU},\mathrm{GDU}\}$. For each $(n,m,t)$ combination, 10 instances were generated, totaling 480 instances.
\end{itemize}

The RRU and RRW types are used in the SMALL instance set to verify the theoretical results from an experimental perspective and to investigate worst-case performance.  
On the other hand, the LARGE instance set with the other types is designed to simulate real-world applications, where the graph is large but customer nodes are geographically concentrated.

\subsection{Computational Results on SMALL Instances}
We performed experiments on all SMALL instances using the exact method, TDA-SA, TDA-NNA and TDA-SPA.

\begin{figure}[!htb]
    \begin{minipage}[b]{0.30\columnwidth}
    \includegraphics[width=\linewidth]{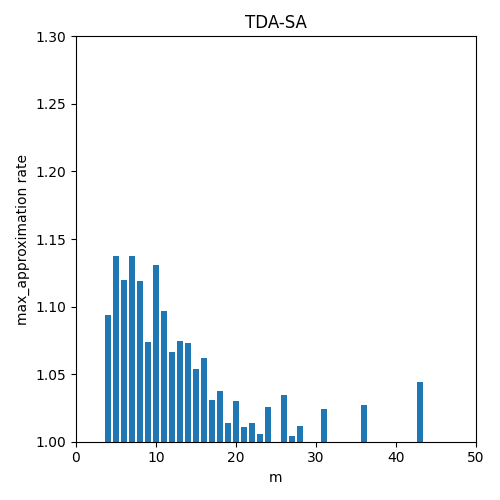}
    \end{minipage}
    \begin{minipage}[b]{0.30\columnwidth}
    \includegraphics[width=\linewidth]{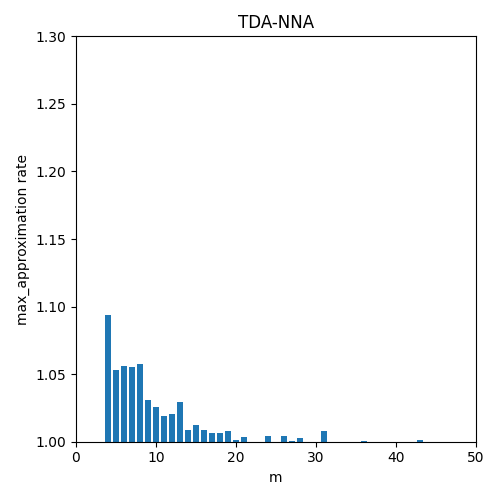}
    \end{minipage}
    \begin{minipage}[b]{0.30\columnwidth}
    \includegraphics[width=\linewidth]{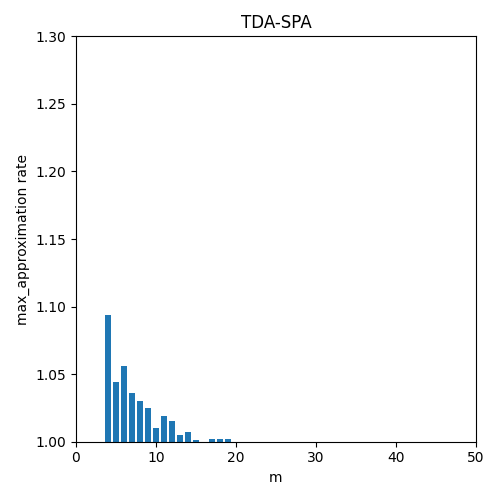}
    \end{minipage}
    \caption{Maximum approximation ratio vs. number of customer nodes $m$ on SMALL instances with $t=\mathrm{RRU}$.}
    \label{fig:maxkinzi_unweight}
\end{figure}

\begin{figure}[!htb]
    \begin{minipage}[b]{0.30\columnwidth}
    \includegraphics[width=\linewidth]{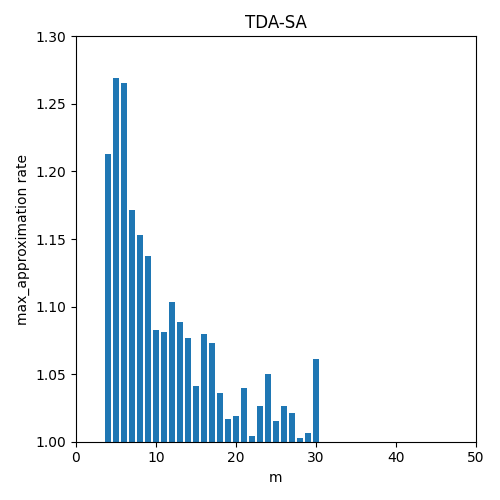}
    \end{minipage}
    \begin{minipage}[b]{0.30\columnwidth}
    \includegraphics[width=\linewidth]{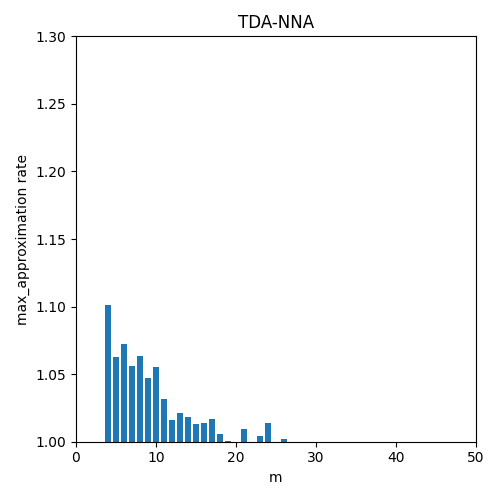}
    \end{minipage}
    \begin{minipage}[b]{0.30\columnwidth}
    \includegraphics[width=\linewidth]{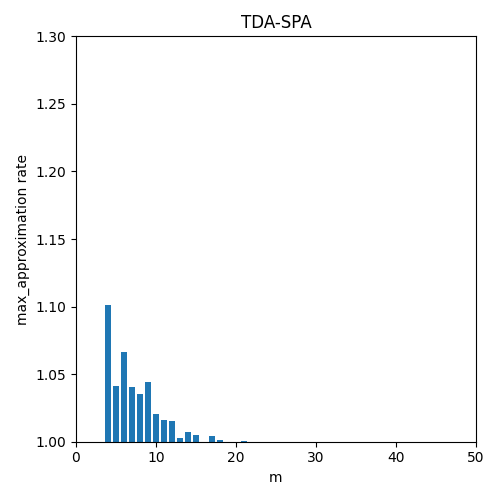}
    \end{minipage}
    \caption{Maximum approximation ratio vs. number of customer nodes $m$ ratio on SMALL instances with $t=\mathrm{RRW}$.}
    \label{fig:maxkinzi_weight}
\end{figure}

The maximum approximation ratios for $\mathrm{RRU}$ and $\mathrm{RRW}$ graphs are shown in Figures~\ref{fig:maxkinzi_unweight} and~\ref{fig:maxkinzi_weight}, respectively.
The horizontal axis represents the number of customer nodes, and the vertical axis shows the maximum approximation ratio over the 50,000 instances.

\begin{figure}[!htb]
    \begin{minipage}[b]{0.30\columnwidth}
    \includegraphics[width=\linewidth]{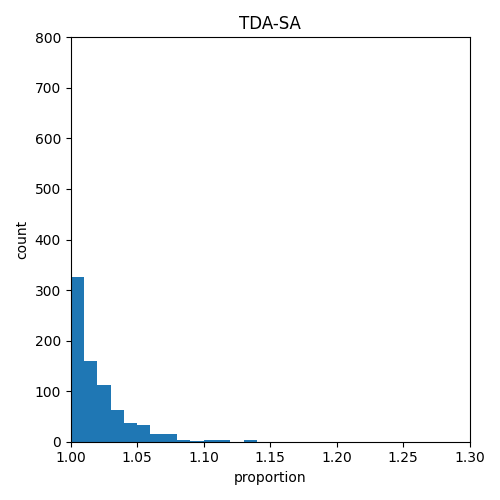}
    \end{minipage}
    \begin{minipage}[b]{0.30\columnwidth}
    \includegraphics[width=\linewidth]{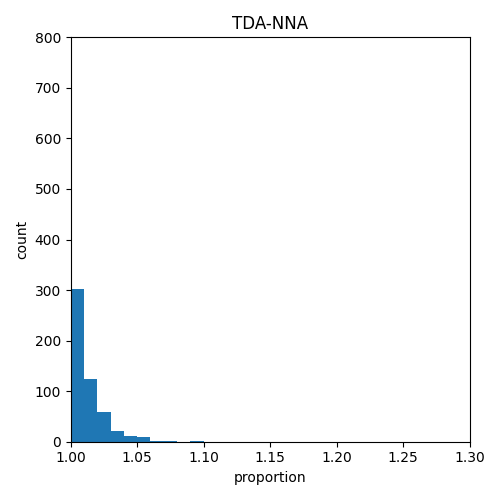}
    \end{minipage}
    \begin{minipage}[b]{0.30\columnwidth}
    \includegraphics[width=\linewidth]{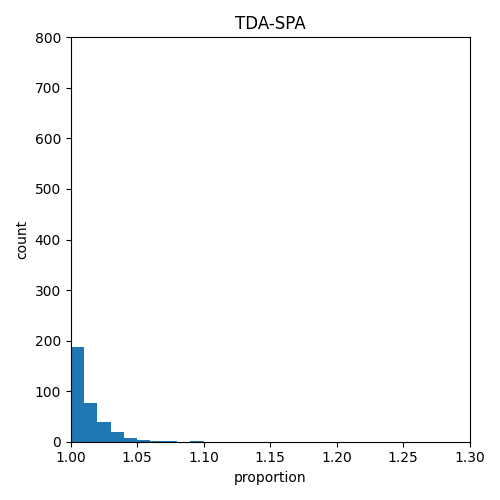}
    \end{minipage}
    \caption{Approximation ratios ($>1$) and their frequencies for SMALL instances with $t=\mathrm{RRU}$.}
    \label{fig:pro_unweight}
\end{figure}

\begin{figure}[!htb]
    \begin{minipage}[b]{0.30\columnwidth}
    \includegraphics[width=\linewidth]{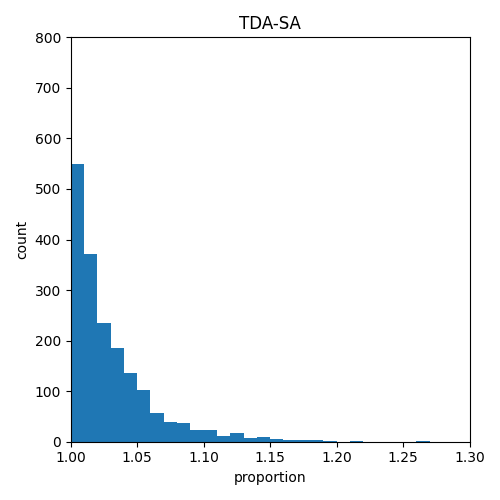}
    \end{minipage}
    \begin{minipage}[b]{0.30\columnwidth}
    \includegraphics[width=\linewidth]{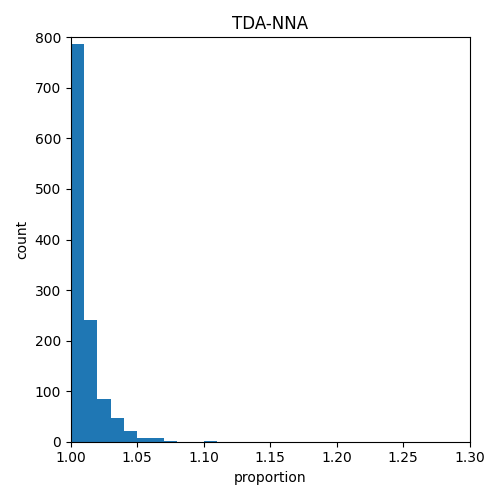}
    \end{minipage}
    \begin{minipage}[b]{0.30\columnwidth}
    \includegraphics[width=\linewidth]{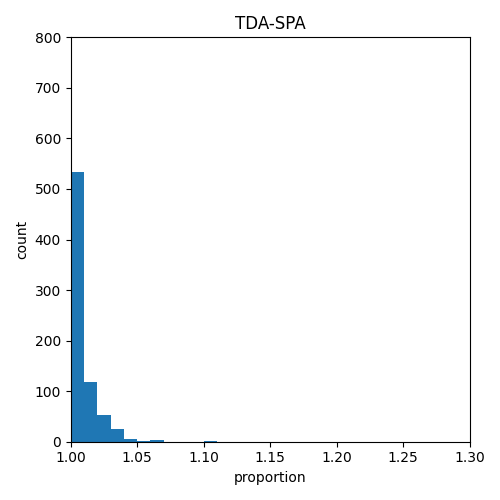}
    \end{minipage}
    \caption{Approximation ratios ($>1$) and their frequencies for SMALL instances with $t=\mathrm{RRW}$.}
    \label{fig:pro_weight}
\end{figure}

Figures~\ref{fig:pro_unweight} and Table~\ref{tab:unweight_count} show the approximation ratios and their frequencies for instances where the proposed methods failed to obtain the optimal solution for $t=\mathrm{RRU}$. The corresponding results for $t=\mathrm{RRW}$ are shown in Figure~\ref{fig:pro_weight} and Table~\ref{tab:weight_count}.

\begin{table}[!htb]
    \centering
    \caption{Proportion of instances where optimal solution was not found and maximum approximation ratio for $t=\mathrm{RRU}$.}
    \label{tab:unweight_count}
    \begin{tabular}{lrr}
        \hline
        & Ratio of suboptimal instances & Max approximation ratio \\
        \hline
        TDA-SA & $775/2500000$ & $1.138$ \\
        TDA-NNA & $534/2500000$ & $1.094$ \\
        TDA-SPA & $340/2500000$ & $1.094$ \\
        \hline
    \end{tabular}
\end{table}

\begin{table}[!htb]
    \centering
    \caption{Proportion of instances where optimal solution was not found and maximum approximation ratio for $t=\mathrm{RRW}$.}
    \label{tab:weight_count}
    \begin{tabular}{lrr}
        \hline
        & Ratio of suboptimal instances & Max approximation ratio \\
        \hline
        TDA-SA & $1822/2500000$ & $1.269$ \\
        TDA-NNA & $1197/2500000$ & $1.105$ \\
        TDA-SPA & $741/2500000$ & $1.101$ \\
        \hline
    \end{tabular}
\end{table}

For TDA-SA, the maximum approximation ratio observed was $1.269$, well below the theoretical upper bound given in Theorem~\ref{theorem:2kinzi}. For TDA-NNA and TDA-SPA, the maximum approximation ratio was $1.101$, which is below the lower bound of $1.2$ shown in Theorem~\ref{theorem:12kinzi}. These results suggest that the theoretical ratio for TDA-NNA and TDA-SPA may be further improved to approach the lower bound from Theorem~\ref{theorem:12kinzi}.

Also, proposed methods were able to obtain the optimal solution for over 99.9\% of the instances. In particular, TDA-NNA and TDA-SPA provided high-quality solutions with approximation ratios below $1.05$ in most of the remaining 0.1\% of instances.

\subsection{Computational Results on LARGE Instances}
The solution quality for LARGE instances showed similar trends to those on SMALL instances. Specifically, all proposed methods obtained an optimal solution for all the tested 480 instances. Tables~\ref{tab:time_RNU}--\ref{tab:time_GDU} summarize the average computational times (in milliseconds) for the exact method, TDA-SA, TDA-NNA and TDA-SPA.

\begin{table}[!htb]
\centering
\caption{Average computational time (ms) for $t=\mathrm{RNU}$.}
\label{tab:time_RNU}
\begin{tabular}{rrrrrrrr}
\hline
$n$ & $m$ & & & Exact method & TDA-SA & TDA-NNA & TDA-SPA \\ \hline
& 2&&& 4  & 1  & 1  & 1 \\ 
$10^4$&8 &&& 17  & 14  & 14  & 14 \\ 
& 32&&& 71  & 76  & 76  & 77 \\ \hline
& 2&&& 77  & 23  & 20  & 24 \\ 
$10^5$&8 &&& 298  & 324  & 331  & 342 \\ 
& 32&&& 1179  & 2389  & 2406  & 2417 \\ \hline 
& 2&&& 1365  & 234  & 186 & 258 \\ 
$10^6$& 8&&& 5463  & 3048  &  3032 & 3249 \\ 
& 32&&& 21817  & 31034 & 31368 & 31569   \\ \hline
& 2&&& 19524  & 4352 & 3502 & 4754   \\ 
$10^7$&8 &&& 78010  & 48673  & 48556 & 51045  \\ 
& 32&&& 312743  & 448818  & 448473 & 451185  \\ \hline

\end{tabular}
\end{table}
\begin{table}[!htb]
\centering
\caption{Average computational time (ms) for $t=\mathrm{RDU}$.}
\label{tab:time_RDU}
\begin{tabular}{rrrrrrrr}
\hline
$n$ & $m$ & & & Exact method & TDA-SA & TDA-NNA & TDA-SPA \\ \hline
& 2&&& 4  & 0  & 0  & 0 \\ 
$10^4$& 8&&& 17  & 2  & 2  & 2 \\ 
& 32&&& 70  & 30  & 30  & 36 \\ \hline
& 2&&& 76  & 0  & 0  & 0 \\ 
$10^5$& 8&&& 298  & 6  & 4  & 7 \\ 
& 32&&& 1179  & 340  & 318  & 397 \\ \hline
& 2&&& 1361  & 20  & 11  & 22 \\ 
$10^6$& 8&&& 5462  & 50  & 25 & 55  \\ 
& 32&&& 21855  & 435  & 263  & 586 \\ \hline
&2 &&& 19542  & 8  & 4 & 9 \\ 
$10^7$& 8&&& 78120  & 1222  & 670 & 1320  \\ 
& 32&&& 312908  & 1889  & 1036  & 2346 \\ \hline
\end{tabular}

\end{table}
\begin{table}[!htb]

\centering
\caption{Average computational time (ms) for $t=\mathrm{GNU}$.}
\label{tab:time_GNU}
\begin{tabular}{rrrrrrrr}
\hline
$n$ & $m$ & & & Exact method & TDA-SA & TDA-NNA & TDA-SPA \\ \hline
& 2&&& 1  & 0  & 0  & 0 \\ 
$10^4$& 8&&& 7  & 0  & 0  & 0 \\ 
& 32&&& 30  & 0  & 0  & 1 \\ \hline
& 2&&& 26  & 0  & 0  & 0 \\ 
$10^5$& 8&&& 100  & 0  & 0  & 0 \\ 
& 32&&& 399  & 0  & 0  & 1 \\ \hline
& 2&&& 351  & 0  & 0  & 0 \\ 
$10^6$& 8&&& 1428  & 0  & 0  & 0 \\ 
& 32&&& 5941  & 0  & 0  & 1 \\ \hline
& 2&&& 4692  & 0  & 0  & 0 \\ 
$10^7$& 8&&& 18082  & 0  & 0  & 0 \\ 
& 32&&& 76030  & 1  & 0  & 1 \\ \hline
\end{tabular}
\end{table}
\begin{table}[!htb]
\centering
\caption{Average computational time (ms) for $t=\mathrm{GDU}$.}
\label{tab:time_GDU}
\begin{tabular}{rrrrrrrr}
\hline
$n$ & $m$ & & & Exact method & TDA-SA & TDA-NNA & TDA-SPA \\ \hline
& 2&&& 1  & 0  & 0  & 0 \\ 
$10^4$& 8&&& 7  & 0  & 0  & 0 \\ 
& 32&&& 30  & 0  & 0  & 1 \\ \hline
& 2&&& 26  & 0  & 0  & 0 \\ 
$10^5$& 8&&& 95  & 0  & 0  & 0 \\ 
& 32&&& 397  & 0  & 0  & 1 \\ \hline
& 2&&& 342  & 0  & 0  & 0 \\ 
$10^6$&8 &&& 1395  & 0  & 0  & 0 \\ 
& 32&&& 5675  & 0  & 0  & 1 \\ \hline
& 2&&& 4353  & 0  & 0  & 0 \\ 
$10^7$&8 &&& 18156  & 0  & 0  & 0 \\ 
& 32&&& 70731  & 0  & 0  & 1 \\ \hline
\end{tabular}
\end{table}

The proposed methods exhibited excellent computational efficiency, across all graph types. Notably, for grid graphs with $10^7$ nodes, all optimal solutions were obtained in less than 1 millisecond.
The slightly increased computation time observed in the RNU case is attributed to the wider search range of the Dijkstra procedures compared to instances of other types with the same number of nodes.

\section{Conclusion}
In this study, we addressed the 1-median problem on large-scale graphs with millions of nodes (potential facility locations), where the number of customer nodes is relatively small and they are spatially concentrated.
We proposed three approximation algorithms: 
TDA-SA, TDA-NNA and TDA-SPA, derived by early termination of Dijkstra's search. Among them, we established approximation guarantees: TDA-NNA and TDA-SPA achieve a ratio of 1.618, with a proven lower bound of 1.2.

Through extensive computational experiments, we confirmed the computational efficiency of the proposed methods compared to the naive exact approach. In terms of solution accuracy, all proposed methods produced optimal solutions for 99.9\% of the tested instances.
Regarding computation time, the proposed methods significantly outperformed the exact method, especially when customer nodes were spatially concentrated, achieving substantial speed-ups.
For the tested grid graphs with 10 million nodes, the proposed algorithms obtained all optimal solutions within 1 millisecond, outperforming the baseline exact method which required over 70 seconds on average. 
Moreover, the experimental results suggest the theoretical approximation ratio may be tightened to 1.2, indicating potential for future theoretical developments.

As a future direction, we aim to explore new exact algorithms that reduce computational complexity by leveraging the obtained theoretical approximation bounds, with the aim of improving upon the current exact method.



\end{document}